\documentclass[11pt,fleqn]{article}
\usepackage[utf8x]{inputenc}
\usepackage[T1]{fontenc}

\usepackage[top=0.75in, bottom=0.75in, left=0.75in, right=0.75in]{geometry}

\usepackage[small]{caption}
\usepackage[noblocks]{authblk}

\usepackage{graphicx}
\usepackage{amsmath}
\usepackage{slashed}
\usepackage{amsfonts}
\usepackage{amssymb}
\usepackage{multirow}
\usepackage{fancyhdr}
\usepackage{cancel}
\usepackage{subcaption}
\usepackage{url}
\usepackage{soul}
\usepackage{calrsfs}
\usepackage{makeidx}
\usepackage{indentfirst}
\usepackage{ragged2e}
\usepackage[dvipsnames]{xcolor}
\usepackage{hyperref}
\usepackage{tikz}
\usepackage[compat=1.1.0]{tikz-feynman}
\usetikzlibrary{shapes,backgrounds,calc,snakes,shadows}
\usepackage[export]{adjustbox}
\usepackage{longtable}

\usepackage{cite}
\usepackage{xcolor,colortbl}

\usepackage[dvipsnames]{xcolor}

\usepackage{multirow} 
\usepackage{hhline}
\usepackage{array}
\newcolumntype{?}{!{\vrule width 1pt}}
\makeatletter
\def\thickhline{%
  \noalign{\ifnum0=`}\fi\hrule \@height \thickarrayrulewidth \futurelet
   \reserved@a\@xthickhline}
\def\@xthickhline{\ifx\reserved@a\thickhline
               \vskip\doublerulesep
               \vskip-\thickarrayrulewidth
             \fi
      \ifnum0=`{\fi}}
\makeatother

\newlength{\thickarrayrulewidth}
\begin{document}
	\title{{\sf Applications of Diagrammatic Renormalization Methods in QCD Sum-Rules }}
	\author[1]{T. de Oliveira\thanks{tdo842@usask.ca}}
	\author[1]{D. Harnett\thanks{derek.harnettt@shaw.ca}}
	\author[2]{A. Palameta\thanks{alexander.palameta@ufv.ca}}
\author[1]{T.G. Steele\thanks{tom.steele@usask.ca}}

\affil[1]{Department of Physics \&
Engineering Physics, University of Saskatchewan, SK, S7N 5E2, Canada}
\affil[2]{Department of Physics, University of the Fraser Valley, Abbotsford, BC, V2S 7M8, Canada}

\maketitle

\begin{abstract}
In QCD sum-rule methods, the fundamental field-theoretical quantities are correlation functions of composite operators that serve as hadronic interpolating fields. One of the challenges of loop corrections to QCD correlation functions in conventional approaches is the renormalization-induced mixing of composite operators. This involves a multi-step process of first renormalizing the operators, and then calculating the correlation functions in this mixed basis. This process becomes increasingly complicated as the number of operators mixed under renormalization increases, 
a situation that is exacerbated as the operator mass dimension increases in important physical systems such as tetraquarks, pentaquarks, and hybrids. 
Diagrammatic renormalization provides an alternative  to  the conventional operator  renormalization approach.  Diagrammatic renormalization methods are outlined
and applied to a variety of QCD sum-rule examples of increasing complexity. 
The results are benchmarked, and the diagrammatic method is contrasted with  
the conventional operator mixing approach. Advantages and conceptual interpretations of the diagrammatic renormalization approach are outlined and technical subtleties are explored.
\end{abstract}
	
\section{Introduction}	

\noindent
QCD sum-rules are founded on the concept of quark-hadron duality where  QCD composite operators, with appropriate quantum numbers and valence content,  are used as interpolating fields to relate hadronic properties to QCD correlation functions of composite operators~\cite{Shifman:1978bx,Shifman:1978by} (see also QCD sum-rule reviews in Refs.~\cite{Reinders:1984sr,narison,Gubler:2018ctz,Albuquerque:2018jkn}).  From this basic formulation, it is immediately evident that  renormalization of the underlying composite operators in the correlation function is a necessary aspect of  QCD sum-rule calculations.   
In most cases, the composite operator renormalization effects enter at next-to-leading loop order (NLO), 
but in  cases of cross-correlators between different currents (e.g., glueballs and quark-antiquark mesons \cite{Narison:1984bv,Harnett:2008cw}), composite operator renormalization enters  at leading-order (LO). 
Thus, in the conventional approach of renormalization constants (or counter-terms), renormalization  of QCD correlation functions requires two distinct steps:
first, replacement of bare quantities (couplings, masses, and fields) 
with their renormalized counterparts and second,
the renormalization of the composite operators themselves. 
Both steps of this conventional renormalization approach must be implemented correctly to cancel the divergences within QCD  
correlation functions used to predict hadronic properties.

Renormalization of composite operators in gauge theories has a long history 
\cite{Dixon:1974ss,Kluberg-Stern:1974iel,Joglekar:1975nu,Deans:1978wn}. 
Under renormalization,  a gauge-invariant operator not only mixes with other gauge-invariant 
operators of equal or lower dimension, but also with equations of motion and non-physical operators \cite{Dixon:1974ss,Kluberg-Stern:1974iel,Joglekar:1975nu} (see also \cite{collins,pascualandtarrach} 
for reviews).    
Although background field methods simplify some aspects of composite operator renormalization \cite{Kluberg-Stern:1975ebk,Tarrach:1981bi}, 
the underlying mixing becomes increasingly complicated as the operator mass dimension increases. 
For example, the renormalization of the simplest single-flavour dimension-six $\bar q q\bar q q$ 
light quark operators involves the mixing of 10 operators \cite{Narison:1983kn}, and the
renormalization of dimension-five mixed quark/gluon $\bar q Gq$ operators mixes 
five or six operators~\cite{Narison:1983kn,Jin:2000ek}. 
Some of these mixings result in ``nuisance operators''  that do not ultimately contribute to correlation functions \cite{Jin:2000ek}, adding extraneous calculational overhead
to sum-rule applications.

The increasing complexity of the conventional approach to renormalization of higher-dimension composite operators provides significant challenges for QCD sum-rule calculations 
nowadays 
with numerous experimental discoveries of exotic hadrons such as multiquark states  (e.g., tetraquarks and pentaquarks)  and  hybrid  (conventional hadrons with valence gluonic content) candidates (see e.g., Refs.~\cite{Brambilla:2019esw,Liu:2019zoy,Meyer:2015eta} for reviews). 
 Next-to-leading order (NLO) QCD sum-rule studies of tetraquark and pentaquark states would require renormalization of dimension-six and dimension-15/2 operators containing heavy quarks for a variety of quantum numbers, flavour content, and colour structures.  It is, therefore, not surprising that
the overwhelming majority of QCD sum-rule studies of heavy-quark tetraquark and pentaquark systems are LO calculations (see e.g., \cite{Liu:2019zoy,Albuquerque:2018jkn} for reviews). To our knowledge, renormalization of the composite operators in these systems has never been studied. 
Thus, there is strong motivation
to develop more efficient composite operator renormalization methodologies for QCD sum-rule applications.

As outlined above, the conventional approach to renormalization is based on Lagrangian counterterms
and extends to the renormalization of composite operators in a natural way (see e.g., Ref.~\cite{pascualandtarrach}).  
However,  the 
diagrammatic renormalization method is a lesser-utilized approach to renormalization 
that is based on individual Feynman diagrams of Lagrangian fields~\cite{bogoliubov,Hepp:1966eg,Zimmermann:1969jj,Collins:1974da} 
(see Refs.~\cite{collins,muta} for an overview).  
The  diagrammatic method can be extended to Feynman diagrams that contain composite operator insertions (see e.g., Ref.~\cite{collins}), and so
is relevant to QCD correlation functions of composite operators occurring in QCD sum-rule methods.

The purpose of this paper is to outline
and apply diagrammatic renormalization methods 
for QCD correlation functions in a variety of examples including $\bar q q$ quark mesons, 
$\bar q q$-glueball mixing, and heavy-light diquarks.
In these examples, the results of diagrammatic renormalization will be shown to agree with conventional 
renormalization approaches, demonstrating validity of the diagrammatic method in QCD sum-rule applications.  
Conceptual insights that emerge from comparing the two renormalization approaches will be emphasized, 
including  techniques for the anomalous dimension factors in the diagrammatic approach.
Technical subtleties that could lead to erroneous results in the diagrammatic technique will be 
highlighted, and the effects of different gauge parameter choices will be outlined. 
Advantages of the diagrammatic approach will be discussed with an emphasis on future applications 
to multiquark systems.

\section{Diagrammatic Renormalization of QCD Correlation Functions}
\label{diag_method_sec}

\noindent
Many QCD sum-rules analyses are based on 
two-point correlation functions of composite operators (currents) $J_{\Gamma_i}(x)$,
\begin{equation}
\Pi_{\Gamma_1\Gamma_2}\left(q\right)=i\int d^Dx \,e^{i q\cdot x}
\langle O\vert T\left[J_{\Gamma_1}(x) J^\dagger_{\Gamma_2}(0) \right]\vert O\rangle \,,
\label{corr_defn}
\end{equation}
where $D=4+2\epsilon$ represents the spacetime dimension for dimensional regularization and $\Gamma_i$ represents the collective quantum numbers associated  with the currents. 
In many cases, the correlation function is diagonal with $\Gamma_1=\Gamma_2$, 
but  non-diagonal correlation functions with $\Gamma_1\ne \Gamma_2$ occur when studying mixed interpretations 
of hadrons (e.g., mixing of glueballs and quark-antiquark mesons).
Generally, some projection operator is applied to~(\ref{corr_defn}) yielding a scalar function
$\Pi(Q^2)$ where $Q^2=-q^2$. (See~(\ref{Pi_v}) for an example.)
Based on its high-energy behaviour,  
$\Pi(Q^2)$
satisfies a dispersion relation,
\begin{equation}
\Pi\left(Q^2\right)=\Pi(0)+Q^2\Pi'(0)+\frac{1}{2}Q^4\Pi''(0)
+\ldots +\frac{1}{n}Q^{2n}\Pi^{(n)}(0)
+Q^{2n+2}\frac{1}{\pi}
\int\limits_{t_0}^\infty
\,\frac{ {\rm Im}\Pi(t)}{t^{n+1}\left(t+Q^2\right)}dt \,,
\label{GenDispRel}
\end{equation}
where the $\Pi^{(n)}(0)$ are  subtraction constants and 
$t_0$ is a field-theoretical threshold.
Except for cases where a low-energy theorem exists (see e.g., Ref.~\cite{Novikov:1981xi}), 
the subtraction constants are unknown and will typically be  divergent.  
However, the divergent subtraction constants are eliminated in QCD sum-rule methods 
(e.g.,  by taking a sufficient number of $Q^2$ derivatives 
or applying some integral 
transform)~\cite{Shifman:1978bx,Shifman:1978by,Reinders:1984sr,narison,Gubler:2018ctz},
and hence, they do not affect physical predictions.  
From a field-theoretical perspective, the $\Pi^{(n)}(0)$ are local divergences because of their $Q^2$  
polynomial structure, a property that becomes relevant in diagrammatic renormalization.  

In conventional renormalization, a renormalized composite operator 
(i.e., current) $J_R(x)$ is expressed as a sum of  
renormalization constants and bare operators
\begin{equation}
  J_R=\sum_{i=1}^n Z_i J^{(i)}_B\,,
\label{operator_mix}
\end{equation}
and hence, in general, there will be renormalization-induced mixing of $n$ composite operators.
Calculation of the $Z_i$ first involves developing 
the operator basis from the general 
principles~\cite{Dixon:1974ss,Kluberg-Stern:1974iel,Joglekar:1975nu,collins,pascualandtarrach} and then calculating Green functions containing the bare operator $J_B^{(i)}$ and QCD fields to determine the renormalization factors  $Z_i$.  
Although there are different approaches to help disentangle the contributions from different $J_B^{(i)}$ (see e.g., Refs.~\cite{Narison:1983kn,Jin:2000ek,Bagan:1989vm}), one must
still compute multiple Green functions to fully determine \eqref{operator_mix}.  
Calculation of the renormalized correlator then proceeds via the renormalized currents \eqref{operator_mix} which, in principle, involves all possible combinations of
currents $J^{(i)}_BJ^{(j)}_B$ and the replacement of the bare coupling/masses with their renormalized versions. 
If the entire renormalization procedure is implemented correctly, then all non-local divergences will cancel, and only local (subtraction term) divergences will remain.  
Thus, in a basis of $n$ mixed bare operators, one anticipates the need to calculate 
at least $n$ Green functions for renormalization of the composite operator and then $n(n+1)/2$ bare-operator correlators, with potentially multiple Feynman diagrams in each case.  
Thus, the calculation of QCD correlation functions becomes increasingly demanding in conventional renormalization as $n$ increases (e.g., $n\sim 10$ for tetraquark systems).
Furthermore, there are very few intermediate benchmarks to ensure accuracy in the final result.
These challenges of computational efficiency and accuracy in conventional renormalization
are the motivation for developing diagrammatic renormalization methods for QCD correlation functions.

The essential idea of diagrammatic renormalization is to renormalize each
Feynman diagram $G$ that occurs in the perturbative expansion of bare quantities through a
subtraction process.
Following Ref.~\cite{collins}, the renormalized diagram $R(G)$ is obtained by first removing 
all (non-local) subdivergences to construct $\bar R(G)$, 
and then applying a counterterm $C(G)$ to remove any remaining 
local divergences
\begin{equation}
R(G)=\bar R(G) + C(G)   \,. 
\end{equation}
In the case of QCD correlation functions, the local divergences 
correspond to the subtraction constants which are already eliminated when forming QCD sum-rules,
so the process is slightly simpler: for a correlation function diagram $G$, 
it is only necessary to construct $\bar R(G)$  by 
recursively subtracting from the bare diagram $U(G)$ 
subdivergences occurring in subdiagrams $\gamma$
\begin{equation}
\bar R(G)=U(G)-\sum_\gamma C_\gamma(G)\,,~C_\gamma (G)={\cal T}\circ \bar R(\gamma)
\label{diag_renorm_def}
\end{equation}
where $ {\cal T}$ isolates the divergent part of the subdiagram $\gamma$ 
(e.g., $\epsilon$-expansion terms that diverge as
$\epsilon\to 0$ in minimal-subtraction schemes). 
Thus, in~\eqref{diag_renorm_def},  
each subdiagram is replaced with a counterterm diagram that subtracts its divergent part.   
After every diagram has been renormalized via~\eqref{diag_renorm_def}, 
the resulting sum of diagrams is 
the renormalized correlation function  
with all mass/coupling parameters understood as their renormalized versions
at the renormalization scale $\nu$ associated with the subtraction scheme ${\cal T}$ 
(typically MS or $\overline{\rm MS}$ scheme).  
The entire process of 
renormalization-induced operator mixing in the conventional renormalization approach 
is obviated, leading to considerable increases in computational efficiency because it is only necessary to compute and renormalize the diagrams associated with the bare correlator.  
Because the process~\eqref{diag_renorm_def} must cancel all non-local divergences for each diagram,
diagrammatic renormalization has a built-in internal diagnostic to improve calculational accuracy.
These features will be illustrated in the applications presented 
in Section~\ref{example_section}. 

\section{QCD Sum-Rule Applications of Diagrammatic Renormalization}
\label{example_section}

\noindent
In this section, NLO (i.e., two-loop) applications of diagrammatic renormalization 
to QCD correlation functions are presented and contrasted with conventional renormalization.
When relevant, calculations will be presented for the (momentum-space) gluon propagator
$ i \delta^{ab}
 \left[ -\frac{g_{\mu\nu}}{k^2}+(1-\xi)\frac{k_\mu k_\nu}{k^4} \right]$
in both Landau ($\xi=0$) and Feynman ($\xi=1$) gauge to provide greater insight and guide gauge
parameter selections in future applications.  
All calculations 
are performed in dimensional regularization with $D=4+2\epsilon$ 
and results are presented in the $\overline{\rm MS}$ scheme
using FeynCalc \cite{Mertig:1990an,Shtabovenko:2016sxi,Shtabovenko:2020gxv},  
TARCER \cite{Mertig:1998vk} implementation of recursion relations for two-loop
integrals \cite{Tarasov:1996br,Tarasov:1997kx}, 
Package-X \cite{Patel:2015},  results for master integrals \cite{Boos:1990rg,Davydychev:1990cq,Broadhurst:1993mw}, 
and HypExp \cite{Huber:2005yg,Huber:2007dx} with HPL \cite{Maitre:2005uu} for the expansion of Hypergeometric functions. 

\subsection{Vector Mesonic Correlation Function for Light Quarks}
\label{vec_corr_sec}

\noindent
The QCD correlation function of light quark vector (mesonic) currents is the most familiar example of a QCD correlation function
\begin{gather}
\Pi^{\mu\nu}(q)=i\int d^D x\,e^{iq\cdot x} 
\langle O\vert T\left[J^\mu(x) J^\nu(0) \right]\vert O\rangle\,,
\label{vec_corr_fn}
\\
J^\mu(x)=\bar\psi(x)\gamma^\mu\psi(x)\,.
\label{vec_current}
\end{gather}
Because the current is conserved, the correlation function \eqref{vec_corr_fn} can be expressed via a single form factor
\begin{gather}
\Pi^{\mu\nu}(q)=\left(g^{\mu\nu}-q^\mu q^\nu/q^2\right)\Pi_v(Q^2)\,,~~Q^2=-q^2
\label{Pi_tensor}
\\
\Pi_v(Q^2)=\frac{1}{\left(D-1\right)}\Pi^\mu_\mu(q)\,.
\label{Pi_v}
\end{gather}
It is well known that the vector current \eqref{vec_current} does not require any renormalization factors \cite{Preparata:1968ioj}
(see also \cite{pascualandtarrach}) so that 
\begin{equation}
   J^\mu_R=     J^\mu_B
    \label{vec_curr_renorm}
\end{equation}
[i.e., $Z=1$ in \eqref{operator_mix}], and hence, the vector current is the 
simplest example to illustrate diagrammatic renormalization.

\begin{figure}[htb]
	\centering	
	\includegraphics{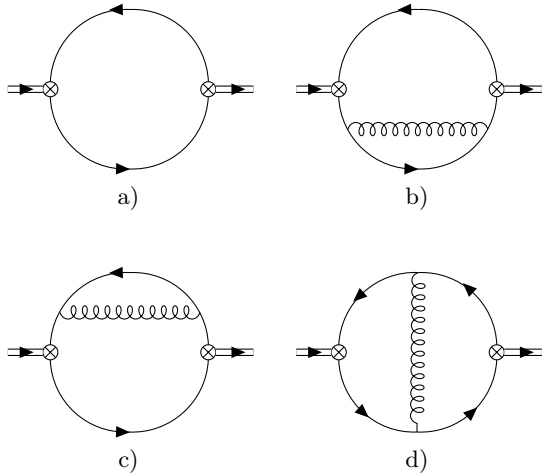}
		\caption{Leading-order Diagram $a$) and next-to-leading order 
(Diagrams $b$--$d$)
	Feynman diagrams for correlation functions of vector and scalar currents. The $\otimes$ denotes either the vector or scalar composite operator Feynman rule and the double line represents the 
	external momentum $q$.  
	Diagrams $b$ and $c$ are the self-energy topology, and Diagram $d$ is the gluon exchange topology.}
	\label{MesonicPerturbativeDiagrams}
\end{figure}

The Feynman diagrams for $\Pi_v$ are shown in Fig.~\ref{MesonicPerturbativeDiagrams} up to NLO
(two-loops), and, in the light-quark chiral limit, 
$\Pi_v$ can be expressed in the form
\begin{equation}
  \Pi_v=\frac{Q^2}{4\pi^2}\left[-L+ \frac{\alpha}{\pi}\left(AL+BL^2 \right)\right]\,,~L=\log\left(Q^2/\nu^2\right)
  \label{vec_corr_form}
\end{equation}
where the first term represents the LO Diagram $a$ of 
Fig.~\ref{MesonicPerturbativeDiagrams}, 
$\nu$ is the $\overline{\rm MS}$ renormalization scale,  
and the NLO Diagrams $b$--$d$
(i.e., those proportional to $\frac{\alpha}{\pi}$)
are parameterized by the coefficients $\{A,B\}$. 
Polynomial terms (i.e., non-logarithmic contributions) are omitted in~\eqref{vec_corr_form} 
because they represent dispersion-relation subtractions that do not contribute to the QCD spectral
function sum-rules. 
The Fig.~\ref{MesonicPerturbativeDiagrams} Feynman Diagram $d$ represents 
a gluon-exchange topology, while 
Diagrams~$b$ and $c$ correspond to the self-energy topology.

The NLO results for each topology in conventional renormalization (which in this case is trivial) are shown in Table~\ref{vec_conv_tab} for the gluon propagator in both Landau and Feynman gauges. 
Although the conventional renormalization process is trivial, 
in Feynman gauge, the individual topologies contain problematic $L/\epsilon$ non-local divergences that only cancel in the total result.  
However, there is considerable simplification in Landau gauge because the quark self-energy is zero in the chiral limit, and  therefore, 
the exchange topology is  free of non-local divergences.
Taking into account normalization conventions, the total NLO result for $\Pi_v$ in Table~\ref{vec_conv_tab} 
agrees with standard calculations (see e.g., Ref.~\cite{pascualandtarrach}) and are gauge-independent 
as required for a gauge-invariant current.

\begin{table}[htb]
    \centering
    \begingroup
    \renewcommand{\arraystretch}{1.5}
    \begin{tabular}{?c?c|c|c?c|c|c?}
    \thickhline
    Gauge & \multicolumn{3}{c?}{Feynman} & \multicolumn{3}{c?}{Landau}\\
    \hline
    Topology & Exchange & Self-Energy & Total & Exchange & Self-Energy & Total\\
    \thickhline
    $A$ & $-\frac{34}{9}+\frac{2}{3}\frac{1}{\epsilon}$ & $\frac{25}{9}-\frac{2}{3}\frac{1}{\epsilon}$ & $-1$ & $-1$ & $0$ & $-1$ \\
    \hline
    $B$ & $\frac{2}{3}$ & $-\frac{2}{3}$ & $0$ & $0$ & $0$ & $0$ \\
    \thickhline
    \end{tabular}
    \endgroup 
    \caption{The NLO conventional renormalization contributions to $\Pi_v$ in~\eqref{vec_corr_form} for
    the Feynman diagrams of Fig.~\ref{MesonicPerturbativeDiagrams} for Feynman gauge ($\xi=1$) and Landau
    gauge ($\xi=0$). The self-energy entry represents the sum of the two diagrams, and hence, 
    the total contribution is the sum of the exchange and self-energy entries.}
    \label{vec_conv_tab}
\end{table}

\begin{figure}[htb]
	\centering
		\includegraphics{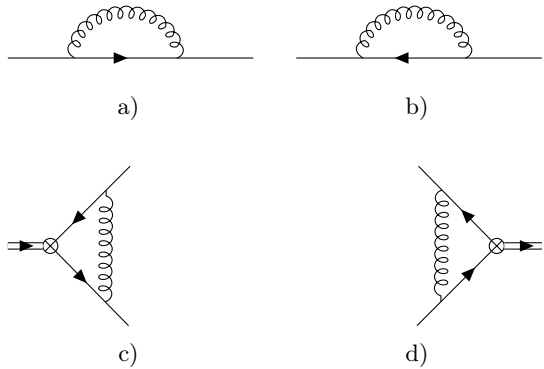}\caption{Subdiagrams extracted from Fig.~\ref{MesonicPerturbativeDiagrams}. 
	Diagrams $a$ and $b$ originate from the NLO self-energy topology 
	and Diagrams $c$ and $d$ from the NLO exchange topology.  
	The $\otimes$ denotes either the vector or scalar composite operator Feynman rule and the double line represents the external momentum $q$.   }
	\label{MesonicSubdiagrams}
\end{figure}

\begin{figure}[htb]
	\centering
		\includegraphics{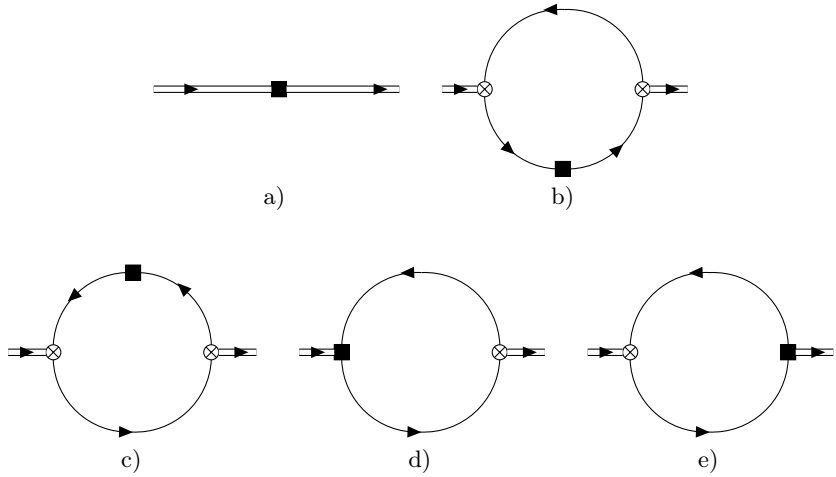}
		\caption{Counterterm diagrams generated by the subdiagrams of Fig.~\ref{MesonicSubdiagrams} 
	and associated with the underlying diagrams in Fig.~\ref{MesonicPerturbativeDiagrams}, where the square $\blacksquare$ denotes the subdivergence insertion,  $\otimes$ denotes either the vector or scalar composite operator Feynman rule, and the double line represents the external momentum $q$. 
	Diagram $a$ is a counterterm  for the LO diagram in Fig.~\ref{MesonicPerturbativeDiagrams}. } 
	\label{MesonicCountertermDiagrams}
\end{figure}

Following the general process outlined in Section~\ref{diag_method_sec}, diagrammatic renormalization for the 
light quark vector correlation function first requires isolation of the subdivergences arising from the
one-loop subdiagram topologies shown in Fig.~\ref{MesonicSubdiagrams}.  
The resulting divergent parts of the subdiagrams, referenced to the topology of the original diagram, 
are given in Table~\ref{vec_SD_tab}.
Counterterm diagrams of Fig.~\ref{MesonicCountertermDiagrams} are generated from the subdivergences of Fig.~\ref{MesonicSubdiagrams}, and then subtracted
to obtain the renormalized  diagram given in Table~\ref{vec_diag_combined_tab} 
for Feynman gauge and 
for Landau gauge. 
As outlined in Section~\ref{diag_method_sec}, within the renormalized entries, the coupling is
interpreted as $\alpha(\nu)$ in  $\overline{\rm MS}$ scheme.
Note that the counterterm for the LO diagram (Diagram $a$ in Fig.~\ref{MesonicCountertermDiagrams}) 
is purely local and, as discussed above in 
Section~\ref{diag_method_sec}, is therefore ignored because it corresponds to a dispersion 
relation subtraction term that does not enter QCD spectral function sum-rules.
As in conventional renormalization, Table~\ref{vec_diag_combined_tab} exhibits  considerable simplifications in Landau gauge because the quark self-energy is zero in the chiral limit. The agreement between the  total NLO result for $\Pi_v$ in conventional and diagrammatic renormalization  (compare Tables~\ref{vec_conv_tab} and \ref{vec_diag_combined_tab}) provides a valuable benchmark for the application of diagrammatic renormalization methods for QCD sum-rules.

\begin{table}[htb]
    \centering
    \begingroup
    \renewcommand{\arraystretch}{1.5}
    \begin{tabular}{?c?c|c?c|c?}
    \thickhline
    Gauge & \multicolumn{2}{c?}{Feynman} & \multicolumn{2}{c?}{Landau}\\
    \hline
    Topology & Exchange & Self-Energy & Exchange & Self-Energy \\
    \hline
    Subdivergence & $-\gamma^\mu \frac{\alpha}{\pi} \frac{1}{3\epsilon}$ & $-i\slashed{p} \frac{\alpha}{\pi} \frac{1}{3\epsilon}$ & $0$ & $0$ \\
    \thickhline
    \end{tabular}
    \endgroup 
    \caption{
    Divergent parts of the one-loop subdiagrams of Fig.~\ref{MesonicSubdiagrams} for the vector current correlation function in both Feynman and Landau gauge.  
    The subdiagrams are classified by the exchange or self-energy topology of the  Fig.~\ref{MesonicPerturbativeDiagrams} diagram from which they originate. 
    The results for the two exchange subdiagrams and the two self-energy subdiagrams are equal.  
    The index $\mu$ is associated with the vector current Feynman rule and the momentum $p$ represents the quark loop momentum flowing through the self-energy.
    }
    \label{vec_SD_tab}
\end{table}

\begin{table}[htb]
    \centering
    \begingroup
    \renewcommand{\arraystretch}{1.5}
    \resizebox{\textwidth}{!}{%
    \begin{tabular}{?c?c?c|c|>{\columncolor[gray]{0.8}}c?c|c|>{\columncolor[gray]{0.8}}c?c?}
    \thickhline
    \multirow{2}{*}{Gauge} & Topology & \multicolumn{3}{c?}{Exchange} & \multicolumn{3}{c?}{Self-Energy} &  \\
    \hhline{|~|-|-|-|-|-|-|-|-|}
       & Diagram & Bare & Counterterm & Renormalized & Bare & Counterterm & Renormalized & Total \\
    \thickhline
    \multirow{2}{*}{Feynman} & $A$ & $-\frac{34}{9}+\frac{2}{3}\frac{1}{\epsilon}$ & $2\left(-\frac{5}{9}+\frac{1}{3}\frac{1}{\epsilon}\right)$ & $-\frac{8}{3}$ & $\frac{25}{9}-\frac{2}{3}\frac{1}{\epsilon}$ & $\frac{10}{9}-\frac{2}{3}\frac{1}{\epsilon}$ & $\frac{5}{3}$& $-1$ \\
    \hhline{|~|-|-|-|-|-|-|-|-|}
       & $B$   & $\frac{2}{3}$ & $2\left(\frac{1}{6}\right)$ & $\frac{1}{3}$ &  $-\frac{2}{3}$ & $-\frac{1}{3}$ & $-\frac{1}{3}$& 0\\
    \thickhline
    \multirow{2}{*}{Landau} & $A$ & $-1$ & $2\times 0=0$ & $-1$ &  $0$ &  $0$ &  $0$ & $-1$ \\
    \hhline{|~|-|-|-|-|-|-|-|-|}
       & $B$ & $0$ & $2\times 0=0$ & $0$ & $0$ & $0$& $0$ & $0$\\
    \thickhline
    \end{tabular}}
    \endgroup 
    \caption{The NLO diagrammatic renormalization contributions to $\Pi_v$ in~\eqref{vec_corr_form} 
    for the Feynman diagrams of Fig.~\ref{MesonicPerturbativeDiagrams} in Feynman gauge $\xi=1$, 
    then Landau gauge $\xi=0$.  
    The bare entries are repeated from Table~\ref{vec_conv_tab}, and the factor of $2$ in the exchange counterterm
    entries is a result of the two identical counterterm diagrams in Fig.~\ref{MesonicCountertermDiagrams}. 
    The (shaded) renormalized entries are obtained by subtracting the counterterm from the bare result. 
    The self-energy entries also represent the sum of two diagram, and hence, the total contribution 
    is the sum of the renormalized exchange and renormalized self-energy entries (shaded columns).
    }
    \label{vec_diag_combined_tab}
\end{table}

\begin{figure}[htb]
	\centering
		\includegraphics{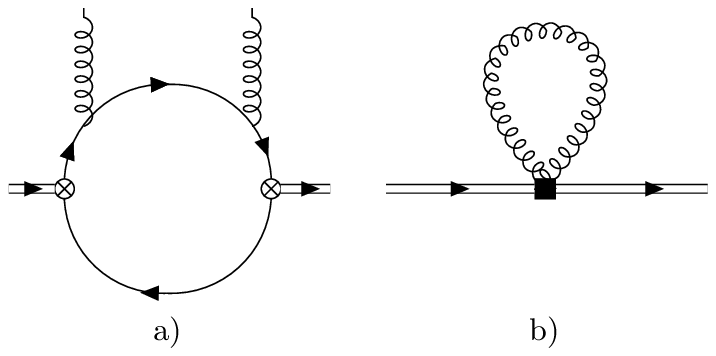}
	\caption{Subdiagram $a$ extracted from the quark loop in the self-energy topologies of
	Fig.~\ref{MesonicPerturbativeDiagrams} and its associated counterterm, i.e.,
	Diagram $b$. 
	The $\otimes$ denotes either the vector or scalar composite operator Feynman rule,  the square
	$\blacksquare$  denotes the subdivergence insertion, 
	and the double line represents the external momentum $q$.}
	\label{additional_subdiagram_fig}
\end{figure}

In principle, there is an additional subdiagram shown in
Fig.~\ref{additional_subdiagram_fig} along with its counterterm diagram.
This counterterm diagram is zero because of the massless tadpole.
However, even if the counterterm diagram was non-zero 
(e.g., the same topology but with a massive line), 
the external momentum would not enter the counterterm diagram,
and hence, the subtraction would correspond to a dispersion-relation subtraction constant 
which does not contribute to QCD sum-rules.

Another technical subtlety in the diagrammatic method that requires attention 
is the role of $D$-dependent factors that occur in the calculation of a subdivergence.
Such $D$-dependent factors could come from contractions of $\gamma$ matrices, for example.
When computing the divergent part of a subdiagram, it is essential to include all
$D$-dependent factors associated with that subdiagram.
Failure to do so will result in errors for the finite parts.  
For example, with computational tools such as 
FeynCalc~\cite{Mertig:1990an,Shtabovenko:2016sxi,Shtabovenko:2020gxv}, 
it may be tempting (for coding simplicity) to defer Dirac/Lorentz algebra until computation of the
counterterm diagram, but this leads to errors in the final result. 
Similarly, it is also important to maintain the ordering of the quark propagators and other Dirac matrices when calculating the subdiagrams (e.g.,  Fig.~\ref{MesonicSubdiagrams} $c$ and $d$).

Despite the simplicity of the vector current renormalization~\eqref{vec_curr_renorm}, 
some of the advantages of diagrammatic renormalization are already evident.  
In particular, no knowledge of the renormalization properties of the vector current were needed in the
diagrammatic approach, an aspect that will be illustrated more powerfully in the subsequent examples where the
renormalization factors are non-trivial.  
Furthermore, the non-local divergences for every diagram must cancel against the 
counterterm diagrams generated from the subdivergences, 
which provides a self-consistency check at the level of each individual diagram. 
In conventional renormalization, divergences only cancel in the sum 
of diagrams, making it more difficult to isolate calculation errors at intermediate stages.

\subsection{Scalar Mesonic Correlation Function for Light Quarks}
\label{scalar_corr_sec}

\noindent
The QCD correlation function of light-quark scalar mesonic currents extends the vector current analysis of Section~\ref{vec_corr_sec} to a situation where the composite operator renormalization is multiplicative but non-trivial ($Z\ne 1$).  
The scalar correlation function is defined as
\begin{gather}
\Pi_s(Q^2)=i\int d^D x\,e^{iq\cdot x} 
\langle O\vert T\left[J(x) J(0) \right]\vert O\rangle \,,~Q^2=-q^2
\label{scalar_corr_fn}
\\
J(x)=\bar\psi(x)\psi(x)\,.
\label{scalar_current}
\end{gather}
Conventional renormalization of the scalar current is given by~\cite{Preparata:1968ioj}
(see also \cite{pascualandtarrach}) 
\begin{equation}
    J_R=     Z_m J_B\,,~Z_m=1+\frac{\alpha}{\pi}\frac{1}{\epsilon}\,,
    \label{scalar_curr_renorm}
\end{equation}
where the notation $Z_m$ is a reminder that the scalar current renormalization constant 
and quark mass renormalization are related.  

The Feynman diagrams for $\Pi_s$ are shown in Fig.~\ref{MesonicPerturbativeDiagrams} up to NLO (two-loops), 
and, in the light-quark chiral limit, $\Pi_s$ can be expressed in the form

\begin{equation}
  \Pi_s=\frac{3Q^2}{8\pi^2}\left[L+ \frac{\alpha}{\pi}\left(AL+BL^2 \right)\right]\,,~L=\log\left(Q^2/\nu^2\right)
  \label{scalar_corr_form}
\end{equation}
where the first term represents the LO diagram of Fig.~\ref{MesonicPerturbativeDiagrams} 
and the NLO corrections are parameterized by the coefficients $\{A,B\}$.   
As before, polynomial terms are omitted in~\eqref{vec_corr_form} because they represent 
dispersion-relation subtractions that do not contribute to the QCD spectral function sum-rules. 

The NLO results for each topology in conventional renormalization are shown in Table~\ref{scalar_conv_tab} 
for the gluon propagator in both Landau and Feynman gauge.  
For conventional renormalization, the factor $Z^2_m$ 
(with $Z_m$ from~\eqref{scalar_curr_renorm}) 
combines with the 
LO diagram to generate the NLO renormalization-induced corrections in Table~\ref{scalar_conv_tab}.

Conventional renormalization is more subtle for the scalar case with cancellations 
of $L/\epsilon$ non-local divergences 
via the renormalization-induced contributions in both Landau and Feynman gauge.
 As in the vector case, there are some simplifications in Landau gauge because the quark self-energy is zero in the chiral limit. 
 Taking into account normalization conventions, the total NLO result for $\Pi_s$ in Table~\ref{scalar_conv_tab}
 agrees with standard calculations (see, e.g.,
 Refs.~\cite{Gorishnii:1990zu,Gorishnii:1991zr,Elias:1998bq}) and are gauge-independent 
 as required for a gauge-invariant current.

\begin{table}[htb]
    \centering
    \begingroup
    \renewcommand{\arraystretch}{1.5}
    \begin{tabular}{?c?c?c|c?c|c?c?}
    \thickhline
    Gauge & & \multicolumn{2}{c?}{Feynman} & \multicolumn{2}{c?}{Landau} & \\
    \hline
    Topology & Renormalization   & Exchange & Self-Energy   & Exchange & Self-Energy  & Total\\
    \thickhline
    $A$ & $-4+\frac{2}{\epsilon}$ & $\frac{38}{3}-\frac{8}{3}\frac{1}{\epsilon}$ & $-3+\frac{2}{3}\frac{1}{\epsilon}$  & $\frac{29}{3}-\frac{2}{\epsilon}$  & $0$ & $\frac{17}{3}$ \\
    \hline
    $B$ & $1$ & $-\frac{8}{3}$ & $\frac{2}{3}$ & $-2$ & $0$ & $-1$ \\
    \thickhline
    \end{tabular}
    \endgroup 
    \caption{The NLO conventional renormalization contributions to $\Pi_s$ 
    in~\eqref{scalar_corr_form} for the Feynman diagrams of Fig.~\ref{MesonicPerturbativeDiagrams}   
    for Feynman gauge ($\xi=1$) and Landau gauge  ($\xi=0$).  
    The self-energy entry represents the sum of the two diagrams,  
    and hence, the total contribution is the sum of the renormalization, exchange, 
    and self-energy entries. 
    The gauge-independent renormalization entry represents the LO diagram combined with the renormalization factor $Z_m^2$ expanded to order $\alpha$. 
    The total result is gauge-independent as expected.     }
    \label{scalar_conv_tab}
\end{table}

Proceeding in the same way as the vector case, diagrammatic renormalization for the 
light quark scalar correlation function begins with isolating the subdivergences arising from 
the one-loop subdiagram topologies shown in Fig.~\ref{MesonicSubdiagrams}.  
The resulting divergent parts of the subdiagrams, referenced to the topology of the original diagram, 
are given in Table~\ref{scalar_SD_tab}. 
Note that the self-energy subdivergence does not depend on the current (compare Tables~\ref{vec_SD_tab} 
and~\ref{scalar_SD_tab}) because the self-energy subdiagram is isolated from the current vertex. 

The Fig.~\ref{MesonicPerturbativeDiagrams} renormalized diagrams obtained by subtracting 
the Fig.~\ref{MesonicCountertermDiagrams} 
 counterterm diagrams generated by the Fig.~\ref{MesonicSubdiagrams} subdivergences (see 
 Table~\ref{scalar_SD_tab}) are given in Table~\ref{scalar_diag_combined_tab} 
 for Feynman gauge and for Landau gauge. 
 As before, the counterterm for the LO diagram (diagram $a$ in Fig.~\ref{MesonicCountertermDiagrams}) 
 is purely local and, therefore, is ignored because it corresponds to a 
 dispersion-relation subtraction term that does not enter QCD spectral function sum-rules. 
As in conventional renormalization, Table~\ref{scalar_diag_combined_tab} exhibits considerable simplifications 
in Landau gauge because the quark self-energy is zero in the chiral limit. 

\begin{table}[htb]
    \centering
    \begingroup
    \renewcommand{\arraystretch}{1.5}
    \begin{tabular}{?c?c|c?c|c?}
    \thickhline
    Gauge & \multicolumn{2}{c?}{Feynman} & \multicolumn{2}{c?}{Landau}\\
    \hline
    Topology & Exchange & Self-Energy & Exchange & Self-Energy \\
    \hline
    Subdivergence & $-\frac{\alpha}{\pi} \frac{4}{3\epsilon}$ & $-i\slashed{p} \frac{\alpha}{\pi} \frac{1}{3\epsilon}$ & $-\frac{\alpha}{\pi} \frac{1}{\epsilon}$ & $0$ \\
    \thickhline
    \end{tabular}
    \endgroup 
    \caption{
    Divergent parts of the  one-loop subdiagrams of Fig.~\ref{MesonicSubdiagrams}  for the scalar current correlation function in both Feynman and Landau gauge.  The subdiagrams are classified by the exchange or self-energy topology of the  Fig.~\ref{MesonicPerturbativeDiagrams} diagram from which they originate. The results for the two exchange subdiagrams and the two self-energy subdiagrams are equal, and the self-energy entries are identical to Table~\ref{vec_SD_tab} because the subdiagram is isolated from the current vertex.  The momentum $p$ represents the quark loop momentum flowing through the self-energy.
    }
    \label{scalar_SD_tab}
\end{table}

\begin{table}[htb]
    \centering
    \begingroup
    \renewcommand{\arraystretch}{1.5}
    \resizebox{\textwidth}{!}{%
    \begin{tabular}{?c?c?c|c|>{\columncolor[gray]{0.8}}c?c|c|>{\columncolor[gray]{0.8}}c?c?}
    \thickhline
    \multirow{2}{*}{Gauge} & Topology & \multicolumn{3}{c?}{Exchange} & \multicolumn{3}{c?}{Self-Energy}  & \\
  \hhline{|~|-|-|-|-|-|-|-|-|}
       & Diagram & Bare & Counterterm & Renormalized & Bare & Counterterm & Renormalized & Total \\
    \thickhline
    \multirow{2}{*}{Feynman} & $A$ & $\frac{38}{3}-\frac{8}{3}\frac{1}{\epsilon}$ & $2\left(\frac{8}{3}-\frac{4}{3}\frac{1}{\epsilon}\right)$ & $\frac{22}{3}$ & $-3+\frac{2}{3}\frac{1}{\epsilon}$ & $-\frac{4}{3}+\frac{2}{3}\frac{1}{\epsilon}$ & $-\frac{5}{3}$& $\frac{17}{3}$ \\
    \hhline{|~|-|-|-|-|-|-|-|-|}
       & $B$   & $-\frac{8}{3}$ & $2\left(-\frac{2}{3}\right)$ & $-\frac{4}{3}$ &  $\frac{2}{3}$ & $\frac{1}{3}$ & $\frac{1}{3}$& $-1$\\
    \thickhline
    \multirow{2}{*}{Landau} & $A$ & $\frac{29}{3}-\frac{2}{\epsilon}$ & $2\left(2-\frac{1}{\epsilon}\right)$  & $\frac{17}{3}$ &  $0$ &  $0$ &  $0$ & $\frac{17}{3}$ \\
    \hhline{|~|-|-|-|-|-|-|-|-|}
       & $B$ & $-2$ & $2\left(-\frac{1}{2}\right)$ & $-1$ & $0$ & $0$& $0$ & $-1$\\
    \thickhline
    \end{tabular}}
    \endgroup 
    \caption{The NLO diagrammatic renormalization contributions to $\Pi_s$ in~\eqref{vec_corr_form} for the bare Feynman diagrams of Fig.~\ref{MesonicPerturbativeDiagrams} and the counterterm diagrams of Fig.~\ref{MesonicCountertermDiagrams} in Feynman gauge $\xi=1$, then Landau gauge $\xi=0$.  
    The bare entries are repeated from Table~\ref{scalar_conv_tab}, and the factor of $2$ in the exchange counterterm entries is a result of the two identical counterterm diagrams in Fig.~\ref{MesonicCountertermDiagrams}. 
    The (shaded) renormalized entries are obtained by subtracting the counterterm from the bare result. 
    The self-energy entries also represent the sum of two diagram, and hence, the total contribution is the sum of the  renormalized exchange and renormalized self-energy entries (shaded columns).
    }
    \label{scalar_diag_combined_tab}
\end{table}

The agreement between the total NLO result for $\Pi_s$ in conventional and diagrammatic renormalization  
(compare Tables~\ref{scalar_conv_tab} and \ref{scalar_diag_combined_tab}) is quite remarkable because the diagrammatic method has not used the Eq.~\eqref{scalar_curr_renorm} renormalization properties of the 
scalar composite operator.  
This provides an important benchmark demonstrating that diagrammatic renormalization methods can calculate QCD correlation functions without any knowledge of the underlying renormalization properties of 
the composite operator.

An important technical subtlety is embedded in the scalar current correlation function.  
Because the coupling in the renormalized correlation function is the 
$\overline{\rm MS}$ scheme coupling $\alpha(\nu)$,
 the non-zero value for $B$ in $\Pi_s$ 
implies that the renormalization-group (RG) equation for the QCD spectral function sum-rule will have 
an anomalous dimension term. 
In the conventional renormalization approach, this is most easily implemented with a quark mass prefactor
in the current~(\ref{scalar_current}) (see, e.g.,~\cite{Elias:1998bq}) 
because $Z_m$ in~\eqref{scalar_curr_renorm} is the quark mass renormalization factor.
Although the diagrammatic method is seemingly oblivious to the renormalization of the scalar operator, its effect implicitly emerges from the diagrammatic calculation, and
the anomalous dimension can be extracted from the correlation function. 
Standard RG methods for QCD sum-rules~\cite{Narison:1981ts} can then be applied after extracting the anomalous dimension from the correlation function.

\subsection{Scalar and Vector Mesonic Correlation Function for Heavy Quarks}
\label{heavy_quark_section}

\noindent
The analysis of vector and scalar mesonic correlation functions will now be extended to massive quarks
with an emphasis on how the diagrammatic renormalization methods are influenced by the quark mass $m$.  
Thus, the focus will be on the divergent parts and the diagrammatic renormalization method, 
and hence, the lengthy expressions for the finite parts will be omitted. 

The inclusion of quark mass does not modify the exchange topology subdivergences
of Fig.~\ref{MesonicSubdiagrams}, so the exchange topology results of Tables~\ref{vec_SD_tab} and \ref{scalar_SD_tab} remain valid. 
As noted above, the self-energy subdivergences for massive quarks given in Table~\ref{mass_SE_SD_tab}
do not depend on the current.

\begin{table}[htb]
    \centering
    \begingroup
    \renewcommand{\arraystretch}{1.5}
    \begin{tabular}{?c?c?c?}
    \thickhline
    Gauge & Feynman & Landau\\
    \hline
    Subdivergence & $-i\left(\slashed{p}-4m \right) \frac{\alpha}{\pi} \frac{1}{3\epsilon}$  & $i m \frac{\alpha}{\pi} \frac{1}{\epsilon}$ \\
    \thickhline
    \end{tabular}
    \endgroup 
    \caption{
    Divergent parts of the heavy-quark, one-loop self-energy subdiagrams of Fig.~\ref{MesonicSubdiagrams}  
    (i.e., subdiagrams originating from a self-energy topology in Fig.~\ref{MesonicPerturbativeDiagrams})  
    in both Feynman and Landau gauge.   
    The results for the two self-energy subdiagrams are equal and are identical for the scalar and vector correlation functions because the self-energy is isolated from the current vertex.  
    The momentum $p$ represents the quark loop momentum flowing through the self-energy.
    }
    \label{mass_SE_SD_tab}
\end{table}

The divergent parts of NLO contributions from Fig.~\ref{MesonicPerturbativeDiagrams}, 
the counterterm diagrams of Fig.~\ref{MesonicCountertermDiagrams} obtained via the Fig.~\ref{MesonicSubdiagrams} subdivergences (Tables~\ref{vec_SD_tab}, \ref{scalar_SD_tab} and   \ref{mass_SE_SD_tab}), and the resulting renormalized diagrams have the general form

\begin{gather}
    \Pi_\Gamma\left(Q^2\right)=
    \frac{m^2}{\pi^2}\frac{\alpha}{\pi}\frac{1}{\epsilon}\left[A+ \frac{B}{\sqrt{w\left(w+1\right)}}\tilde L \right]\,,~ Q^2=-q^2\,,
    \label{mass_corr_form}
    \\
    w=\frac{Q^2}{4m^2}\,,~\tilde L=\log{\left[ \frac{\sqrt{w+1}+\sqrt{w}}{\sqrt{w+1}-\sqrt{w}} \right] }~,
    \label{mass_corr_defns}
\end{gather}
where $A$ and $B$ are polynomials in $w$ and $\Gamma\in\{s,v\}$.
Because we find that all $A$ contributions are local, 
they correspond to a dispersion-relation subtraction that can be 
ignored. Similarly the LO counterterm diagram can be ignored as previously discussed.
However, the $\tilde L$ divergent structure is problematic, and hence, 
$B=0$ must result for the renormalized diagrams.   

The NLO divergent parts of the heavy-quark vector and scalar correlation functions for 
the bare diagrams of Fig.~\ref{MesonicPerturbativeDiagrams}
and the counter-term diagrams of Fig.~\ref{MesonicCountertermDiagrams} 
generated by the Fig.~\ref{MesonicSubdiagrams} subdivergences  (see Tables~\ref{vec_SD_tab}, 
\ref{scalar_SD_tab}, and~\ref{mass_SE_SD_tab}) are given in 
Table~\ref{massive_diag_combined_tab} for Feynman gauge and for Landau gauge. 
Similar to the light-quark analyses, Table~\ref{massive_diag_combined_tab} shows that there are still 
some simplifications in Landau gauge for the heavy-quark vector and scalar correlation functions.

\begin{table}[htb]
    \centering
    \begingroup
    \renewcommand{\arraystretch}{1.5}
    \resizebox{\textwidth}{!}{%
    \begin{tabular}{?c?c?c?c|c?c|c?}
    \thickhline \multirow{2}{*}{Current} &
    \multirow{2}{*}{Gauge} & Topology & \multicolumn{2}{c?}{Exchange} & \multicolumn{2}{c?}{Self-Energy}  \\
    \hhline{|~|~|-|-|-|-|-|}
   &    & Diagram & Bare & Counterterm  & Bare & Counterterm \\
    \thickhline
   \multirow{2}{*}{Vector} & Feynman &$B$ & $\frac{1}{3}\left(w+1\right)\left(2w-1\right) $ & $2\left[\frac{1}{6}\left(w+1\right)\left(2w-1\right)\right] $  
   & $-\frac{1}{6}\left(4w^2+2w+1\right) $ & $-\frac{1}{6}\left(4w^2+2w+1\right) $ 
   \\ 
     \hhline{|~|-|-|-|-|-|-|}
      & Landau &$B$ & $0$ & $0$  & $-\frac{3}{2}$ & $-\frac{3}{2}$  \\ 
    \thickhline
     \multirow{2}{*}{Scalar} & Feynman &$B$ & $-4\left(w+1\right)^2$ & $2\left[-2\left(w+1\right)^2\right]$ & $\frac{1}{2}\left(w+1\right)\left(2w-7\right)$ & $\frac{1}{2}\left(w+1\right)\left(2w-7\right)$\\[2pt]
     \hhline{|~|-|-|-|-|-|-|}
      & Landau &$B$ & $-3\left(w+1\right)^2$ & $2\left[-\frac{3}{2}\left(w+1\right)^2\right]$  & $-\frac{9}{2}\left(w+1\right)$ & $-\frac{9}{2}\left(w+1\right)$   \\[2pt]
    \thickhline
    \end{tabular}}
    \endgroup 
    \caption{The NLO divergent contributions to $\Pi_\Gamma$ in~\eqref{mass_corr_defns} 
    for the bare Feynman diagrams of Fig.~\ref{MesonicPerturbativeDiagrams} 
   and the  counterterm diagrams of Fig.~\ref{MesonicCountertermDiagrams} 
    in Feynman gauge $\xi=1$, then Landau gauge $\xi=0$. 
    The vector current is $\Gamma=v$ and the  scalar  is $\Gamma=s$.
    The factor of $2$ in the exchange counterterm entries is a result of the two identical counterterm diagrams in Fig.~\ref{MesonicCountertermDiagrams}, 
    and the self-energy entries also represent the sum of two diagrams.   Thus, the renormalized diagram is the difference between the bare and 
    counterterm entries, resulting in $B=0$ for the renormalized diagrams.
    }
    \label{massive_diag_combined_tab}
\end{table}

As is evident from Table~\ref{massive_diag_combined_tab}, 
the non-local divergences of each diagram are cancelled by its counterterm diagrams, 
resulting in $B=0$ for the renormalized diagrams as required in the diagrammatic renormalization method.  
Once again, it is remarkable that the diagrammatic renormalization method does not require any knowledge of the conventional renormalization of the underlying vector and scalar composite operators [see Eqs.~\eqref{vec_curr_renorm} and \eqref{scalar_curr_renorm}] in the correlation function.  

\subsection{Heavy-Light Diquark Correlation Functions}
\label{diquark_renorm_section}

\noindent
Heavy-light diquark systems are important within  constituent diquark models of closed charm 
$\bar c c\bar q q$ tetraquark systems (see e.g., Ref.~\cite{Maiani:2004vq}).  
These diquark systems have also been studied in QCD sum-rules at NLO 
 for a variety of quantum numbers~\cite{Kleiv:2013dta}
using conventional renormalization of the diquark composite operators up to two-loop 
order~\cite{Kleiv:2010qk}.  
Correlation functions of heavy-light diquarks  thus provide an interesting 
QCD system for exploring diagrammatic renormalization methods.

The heavy-light diquark correlation function is defined as
\begin{gather}
\Pi^{(\Gamma)}\left(q\right) = i \int d^D x \, e^{i q \cdot x} \langle 0 | T\left[ \right. J^{(\Gamma)}_{\alpha}\left(x\right) S_{\alpha\omega} \left[x\,,0\right] J^{(\Gamma)\dagger}_{\omega}\left(0\right) \left. \right] |0\rangle \,,
\label{diquark_correlation_fn}
\end{gather}
where $\alpha$, $\omega$ are colour indices, $\Gamma$ indicates the quantum numbers, and gauge-invariant information is extracted using the Schwinger string  (see Refs.~\cite{Dosch:1988hu,Jamin:1989hh,Kleiv:2013dta}) 
\begin{gather}
\begin{split}
S_{\alpha\omega} \left[x\,,0\right] = P \; \exp\left[{ig\frac{\lambda^a}{2}\int_0^x dz^\mu\ A^a_\mu\left(z\right)}\right]_{\alpha\omega}
\,,
\label{general_schwinger_string}
\end{split}
\end{gather}
where $P$ is the path ordering operator. 
The heavy-light diquark currents are 
\begin{gather}
J^{(\Gamma)}_\alpha = \epsilon_{\alpha\beta\gamma} Q^{T}_{\beta} C{{O}}_\Gamma q_{\gamma} \,,
\label{diquark_currents}
\end{gather}
where $C$ is the charge conjugation operator, $T$ denotes  transpose, $Q$ is a heavy quark, and $q$ is a light quark~\cite{Dosch:1988hu,Jamin:1989hh}. The operators ${{O}}_\Gamma=\gamma_5\,,I\,,\gamma_\mu\,,\gamma_\mu\gamma_5$ respectively couple to scalar $S$ $\left(J^P=0^+\right)$, pseudoscalar $P$ $\left(0^-\right)$, axial vector $A$  $\left(1^+\right)$, and vector $V$ $\left(1^-\right)$ heavy-light diquarks.  Analogous to \eqref{Pi_tensor}, the axial vector and vector projections of the diquark correlation  functions are given by
\begin{gather}
\Pi^{\rm \left(A,V\right)} \left(Q^2\right) = \frac{1}{D-1}\left(\frac{q^\mu q^\nu}{q^2} - g^{\mu\nu} \right) \Pi^{\rm \left(A,V\right)}_{\mu\nu} \left(q\right) \,.
\label{vector_projection}
\end{gather}
Up to NLO, the explicit cancellation of the gauge parameter via the Schwinger string \eqref{general_schwinger_string} has been demonstrated,
and it has been shown that $S_{\alpha\omega} \left[x\,,0\right]=\delta_{\alpha\omega}$  in Landau gauge
\cite{Dosch:1988hu,Jamin:1989hh,Kleiv:2013dta}.  
Thus, up to NLO, the gauge-invariant information content of the diquark correlation function is extracted in Landau gauge where the Schwinger string becomes the trivial
identity operator in colour space.  
 
Conventional renormalization of the diquark composite operators up to order $\alpha$ is given by \cite{Kleiv:2013dta,Kleiv:2010qk}
\begin{gather}
 \left.J^{\,(\Gamma)}_\alpha\right._R = Z_d^{\,(\Gamma)} \left.J^{\,(\Gamma)}_\alpha\right._B\,; \quad Z_d^{\rm \,(S)} = 1 + \frac{\alpha}{2\pi\epsilon} \,, \quad Z_d^{\rm \,(P)} =  1 + \frac{\alpha}{2\pi\epsilon} \,, \quad Z_d^{\rm \,(A)}=1 \,, \quad Z_d^{\rm\,(V)}=1 \,.
\label{diquark_renorm_factor}
\end{gather}
Thus, in conventional renormalization up to NLO, there will only be renormalization-induced diagrams in the scalar  ($S$) and pseudoscalar ($P$) channels. 
In Ref.~\cite{Kleiv:2013dta}, the diquark correlation functions have been calculated up to NLO 
in conventional renormalization, which provides a detailed benchmark for diagrammatic renormalization.

\begin{figure}[htb]
	\centering
		\includegraphics{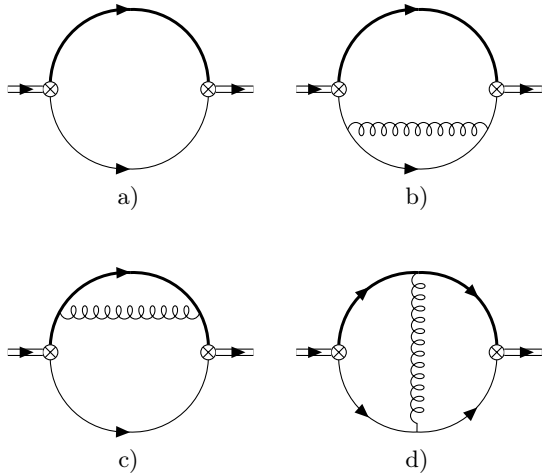}
	\caption{
	Leading-order (Diagram $a$) and  NLO 
	Feynman diagrams for correlation functions of heavy-light diquark currents. 
	The $\otimes$ denotes the diquark composite operator Feynman rule for ${{O}}_\Gamma$, 
	and the double line represents the external momentum $q$. 
	The thick line represents the heavy quark, and the thin line represents the light quark.}
	\label{HLPerturbativeDiagrams}
\end{figure}

\begin{figure}[htb]
	\centering
		\includegraphics{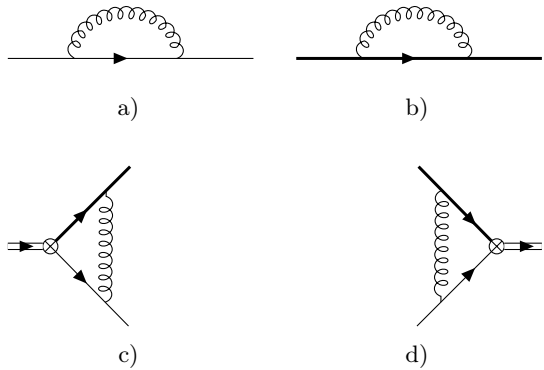}
	\caption{
	Subdiagrams extracted from Fig.~\ref{HLPerturbativeDiagrams}. 
	Diagrams $a$ and $b$ originate from the NLO self-energy topology and 
	Diagrams  $c$ and $d$ from the NLO exchange topology.  
	The $\otimes$ denotes the diquark composite operator Feynman rule for ${{O}}_\Gamma$, 
	and the double line represents the external momentum $q$. 
	The thick line represents the heavy quark, and the thin line represents the light quark. 
	}
	\label{HLSubdiagrams}
\end{figure}

\begin{figure}[htb]
	\centering
		\includegraphics{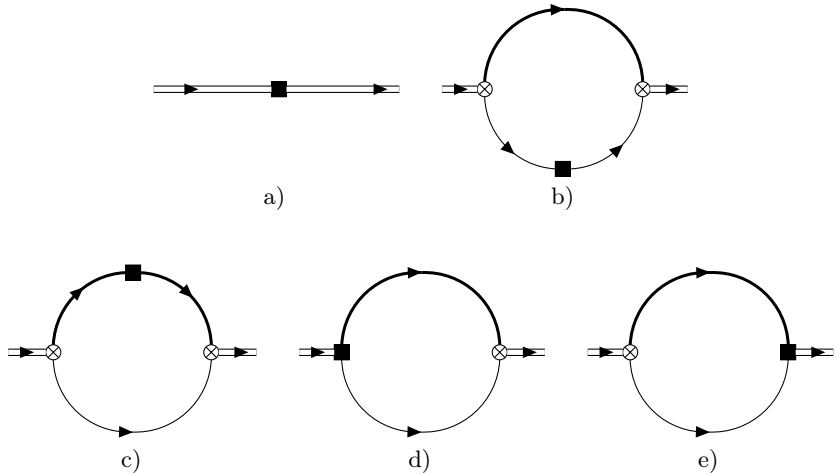}
	\caption{Counterterm diagrams generated by the subdiagrams of Fig.~\ref{HLSubdiagrams} 
	and associated with the underlying diagrams in Fig.~\ref{HLPerturbativeDiagrams}, where the square $\blacksquare$ denotes the subdivergence insertion,  $\otimes$ denotes the diquark composite operator Feynman rule for ${{O}}_\Gamma$, and the double line represents the external momentum $q$. 
	Diagram $a$ is a counterterm for the LO diagram in Fig.~\ref{HLPerturbativeDiagrams}.}
	\label{HLCountertermDiagrams}
\end{figure}

The Feynman diagrams for $\Pi^{(\Gamma)}$ are shown in Fig.~\ref{HLPerturbativeDiagrams} up to NLO
(two-loops), their one-loop subdiagrams are shown in Fig.~\ref{HLSubdiagrams}, 
and the counterterm diagrams generated by the subdiagrams are shown in Fig.~\ref{HLCountertermDiagrams}.
The diagram topologies are classified by light-quark self energy, heavy-quark self-energy, 
and gluon exchange.

The non-local divergences associated with Figs.~\ref{HLPerturbativeDiagrams} and \ref{HLCountertermDiagrams} have the general form 

\begin{equation}
   \Pi^{(\Gamma)}\left(Q^2\right)=B\frac{m^2}{\pi^2}\frac{\alpha}{\pi}\frac{1}{\epsilon}  \frac{\left(1+w \right)}{w^2}\log{(1+w)}\,,~w=\frac{Q^2}{m^2}
   \label{diquark_corr_form}
\end{equation}
where $m$ is the heavy quark mass and $B$ is a polynomial in $w$. The finite parts involve approximately 15 structures including dilogarithms and trilogarithms (see Ref.~\cite{Kleiv:2013dta}) and are not presented for brevity, but are discussed below.

The subdivergences arising from the Fig.~\ref{HLSubdiagrams} subdiagrams are given in Table~\ref{diquark_SD_tab}.  
The non-local divergent parts of the resulting Fig.~\ref{HLCountertermDiagrams} counterterm diagrams, 
the NLO diagrams of Fig.~\ref{HLPerturbativeDiagrams}, 
and the renormalized diagrams are given in Table~\ref{diquark_diag_combined_tab}. 
The LO counterterm diagrams are ignored as discussed previously.

\begin{table}[htb]
    \centering
    \begingroup
    \renewcommand{\arraystretch}{1.5}
    \begin{tabular}{?c?c?c?c?}
    \thickhline
   $J^P$  & $q$ Self-Energy & $Q$ Self-Energy & Exchange\\
    \hline
$0^\pm$ & $0$ & $im\frac{\alpha}{\pi}\frac{1}{\epsilon}$ & $\frac{\alpha}{\pi}\frac{1}{\epsilon}{{O}}_\Gamma$
  \\
  \hline
  $1^\pm$ & $0$ &  $im\frac{\alpha}{\pi}\frac{1}{\epsilon}$ & $0$ \\
    \thickhline
    \end{tabular}
    \endgroup 
    \caption{
    Divergent parts of the  heavy-light diquark one-loop subdiagrams of Fig.~\ref{HLSubdiagrams} 
    in Landau gauge as required for the Schwinger string simplification.   
    The results for the two self-energy subdiagrams are equal to the Landau gauge results in Tables~\ref{vec_SD_tab}, \ref{scalar_SD_tab}, \ref{mass_SE_SD_tab} 
    and are identical for all quantum numbers because the self-energy is 
    isolated from the diquark current vertex.  
    The results for the two exchange diagrams are also equal, and 
    ${{O}}_\Gamma$ is the appropriate operator for the $J^P$ quantum numbers. 
    }
    \label{diquark_SD_tab}
\end{table}

\begin{table}[htb]
    \centering
    \begingroup
    \renewcommand{\arraystretch}{1.5}
    \resizebox{\textwidth}{!}{%
    \begin{tabular}{?c?c?c|c?c|c?c|c?}
    \thickhline \multirow{2}{*}{$J^P$} &
     Topology & \multicolumn{2}{c?}{Exchange} & \multicolumn{2}{c?}{$Q$ Self-Energy} & \multicolumn{2}{c?}{$q$ Self-Energy} \\
    \hhline{|~|-|-|-|-|-|-|-|}
     & Diagram & Bare & Counterterm  & Bare & Counterterm & Bare & Counterterm\\
    \thickhline
   $0^\pm$ &$B$ & $-\frac{3}{4}w\left(w+1\right)$ & $2\left[-\frac{3}{8}w\left(w+1\right)\right] $  
   & $-3w $ & $-3w $ & $0$ & $0$
    \\ 
    \thickhline
    $1^\pm $  &$B$ & $0$ & $0$ & $-\frac{3}{2}\left(w-1\right)$ & $-\frac{3}{2}\left(w-1\right)$
    & $0$ & $0$
    \\ 
    \thickhline
    \end{tabular}}
    \endgroup 
    \caption{The NLO divergent  contributions to $\Pi^{(\Gamma)}$ in \eqref{diquark_corr_form} for the bare Feynman diagrams of Fig.~\ref{HLPerturbativeDiagrams} 
   and the  counterterm diagrams of Fig.~\ref{HLCountertermDiagrams} in Landau gauge as   required for the Schwinger string simplification.
   The factor of $2$ in the exchange counterterm entries is a result of the two identical counterterm diagrams in Fig.~\ref{HLCountertermDiagrams}.  The renormalized diagram is thus the difference between the bare and counterterm entries, resulting in $B=0$ for the renormalized diagrams.
    }
    \label{diquark_diag_combined_tab}
\end{table}

As is evident from Table~\ref{diquark_diag_combined_tab}, 
the non-local divergences of each diagram are cancelled by its counterterm diagram, 
resulting in $B=0$ for the renormalized diagrams as required in the diagrammatic renormalization method.
It has also been verified that the diagrammatically-renormalized finite parts, 
where the coupling and mass are the $\overline{\rm MS}$-scheme quantities 
$\alpha(\nu)$ and $m(\nu)$,
agree with the conventional renormalization results of Ref.~\cite{Kleiv:2013dta}, 
providing a detailed validation of diagrammatic renormalization methods 
in QCD spectral function sum-rules.  
As in all previous examples, it is emphasized that the diagrammatic renormalization method 
does not require any knowledge of conventional renormalization~\eqref{diquark_renorm_factor} 
for the underlying diquark composite operators in the correlation function.  

At this stage, another calculational efficiency of diagrammatic methods becomes 
apparent in the diquark analysis.  
In conventional renormalization, it is somewhat cumbersome to implement mass renormalization 
in the LO diagram, but no comparable challenges exist in the diagrammatic approach.
Furthermore, another diagrammatic self-consistency check becomes evident for the examples developed.
In every case where there are no non-local divergences in the bare diagram, all corresponding subdiagrams are finite (i.e., no subdivergence). 
This property can be used to identify calculation errors within individual diagrams to improve 
accuracy in loop computations. 

\subsection{Scalar Quark Meson Glueball Mixed Correlation Function}
\label{quark_glue_mix_sec}

\noindent
In the diagrammatic renormalization examples presented above, the composite operators 
were multiplicatively renormalizable, 
and the correlation functions were diagonal (i.e., contained a single current).  
The mixed correlator of the scalar light quark meson and scalar glueball operators 
(relevant to mixing of scalar quark mesons with
glueballs~\cite{Narison:1984bv,Harnett:2008cw,Narison:1996fm}) is defined by 
\begin{gather}
  \Pi_{gq}\left(Q^2\right)=i\int\,d^Dx\,e^{i q\cdot x}\left\langle 0
  \vert T\left[ J_g(x)J_q(0)\right] \vert 0 \right\rangle
  ~,~ Q^2\equiv -q^2~,
  \label{mix_corr}
  \\
  J_g(x)=G^a_{\mu\nu}G^a_{\mu\nu}=G^2\,,~J_q(x)=\bar q (x) q(x)
  \label{glue_mix_currrents}
\end{gather}
where  $J_g$ is the scalar glueball current and  $J_q(x)$ is identical to the scalar meson 
current~\eqref{scalar_current}.  
Conventional renormalization of the scalar glueball operator~\cite{Tarrach:1981bi,pascualandtarrach} 
is one of the most familiar cases of operator mixing under renormalization,
\begin{gather}
\left[J_g\right]_R=\left[G^2\right]_R=  \left(1+\frac{\beta_0}{\epsilon}\frac{\alpha}{\pi} \right) \left[G^2\right]_B
-4\frac{\alpha}{\pi}\frac{1}{\epsilon}\left[m_q\overline{q}q\right]_B   \,, 
\label{GG_renorm}
\end{gather}
where $\beta_0=\frac{11}{4}-\frac{1}{6}n_f$ is the one-loop  $\beta$ function coefficient
and, for simplicity, only a single (light) quark flavour of mass $m_q$ has been 
included (extension to additional light flavours is straightforward).
At first order in $\alpha$ and to leading order in the chiral expansion of the light-quark mass $m_q$, 
the general form of $\Pi_{gq}$ is 

\begin{equation}
   \Pi_{gq}=m_q\frac{Q^2}{\pi^2}\frac{\alpha}{\pi} 
   \left[ AL+ BL^2\right]\,,~L=\log{\left(\frac{Q^2}{\nu^2}\right)}\,,
   \label{pi_gq_form}
\end{equation}
and, as before, polynomial terms are omitted in~\eqref{pi_gq_form} 
because they do not contribute to the QCD spectral function sum-rules.

\begin{figure}[htb]
	\centering
		\includegraphics{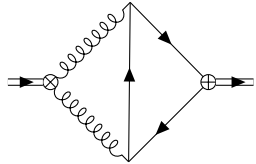}
	\caption{Two-loop diagram for $\Pi_{gq}$ where 
	$\otimes$ represents the Feynman rule for the $J_g$ glueball current 
	and $\oplus$ is the Feynman rule for the $J_q$ quark scalar current.
	}
	\label{Glue_mix_PerturbativeDiagram}
\end{figure}

In conventional renormalization to first-order in $\alpha$, $\Pi_{gq}$ is given by the (two-loop) 
Feynman diagram of Fig.~\ref{Glue_mix_PerturbativeDiagram} combined with the renormalization-induced
Diagram $a$ of Fig.~\ref{MesonicPerturbativeDiagrams}
and the $-4m_q\frac{\alpha}{\pi}\frac{1}{\epsilon}$ prefactor of the scalar quark operator 
$\bar q q$ in~\eqref{GG_renorm}. 
(The $\frac{\beta_0}{\epsilon}\frac{\alpha}{\pi}$ prefactor of the glueball operator 
in~\eqref{GG_renorm} leads to a higher-order $\alpha$ contribution to $\Pi_{gq}$.)
The (gauge-independent) conventional renormalization results given in Table~\ref{glue_mix_conv_tab} 
show that the non-local $\frac{L}{\epsilon}$ divergence from the bare loop is cancelled by the
renormalization-induced diagram.

\begin{table}[htb]
    \centering
    \begingroup
    \renewcommand{\arraystretch}{1.5}
    \begin{tabular}{?c?c?c?c?}
    \thickhline
     & Bare & Renormalization & Total\\
    \hline
$A$ &  $\frac{3}{\epsilon}-\frac{35}{2}$ & $-\frac{3}{\epsilon}+6$ & 
$-\frac{23}{2}$
  \\
  \hline
  $B$ & $3$ &  $-\frac{3}{2}$ & $\frac{3}{2}$ \\
    \thickhline
    \end{tabular}
    \endgroup 
    \caption{
   Conventional renormalization contributions to $\Pi_{gq}$ in~\eqref{pi_gq_form}.  
   The bare entry represents the Feynman diagram of Fig.~\ref{Glue_mix_PerturbativeDiagram}.
   The renormalization entry represents renormalization-induced Diagram $a$ of 
   Fig.~\ref{MesonicPerturbativeDiagrams} 
   (with the scalar quark meson $\otimes$ operator) combined with the renormalization factor 
   $-4m_q\frac{\alpha}{\pi}\frac{1}{\epsilon}$. 
   The total contribution is the sum of the (gauge-independent) bare and renormalization entries.
    }
    \label{glue_mix_conv_tab}
\end{table}

\begin{figure}[htb]
	\centering
		\includegraphics{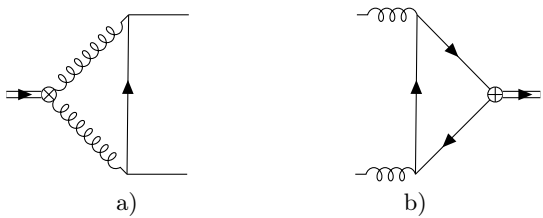}
	\caption{
	Subdiagrams extracted from Fig.~\ref{Glue_mix_PerturbativeDiagram}. 
	Diagram $a$ is the gluon vertex topology and Diagram $b$ is the quark vertex topology. 
	The double line represents the external momentum $q$, $\otimes$ denotes the $G^2$ composite operator
	Feynman rule, and $\oplus$ denotes the $\bar q q$ composite operator Feynman rule. 
	}
	\label{GluonicSubdiagrams}
\end{figure}

\begin{figure}[htb]
	\centering
		\includegraphics{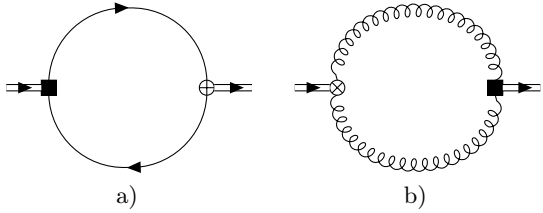}
	\caption{Counterterm diagrams generated by the subdiagrams of Fig.~\ref{GluonicSubdiagrams} 
	and associated with the underlying diagram in Fig.~\ref{Glue_mix_PerturbativeDiagram}, where the square $\blacksquare$ denotes the subdivergence insertion,  $\otimes$ denotes  the $G^2$ composite operator Feynman rule,   $\oplus$ denotes the $\bar q q$ composite operator Feynman rule, and the double line represents the external momentum $q$. 
}
	\label{GluonicCountertermDiagrams}
\end{figure}

In the diagrammatic renormalization method for $\Pi_{gq}$, the one-loop subdiagrams for Fig.~\ref{Glue_mix_PerturbativeDiagram} are shown in Fig.~\ref{GluonicSubdiagrams}.  
The subdiagrams are classified as either a gluon-vertex or quark-vertex topology, 
and their resulting divergent parts 
are given in Table~\ref{glue_mix_subdiag_tab}. 
The results for the bare diagram of Fig.~\ref{Glue_mix_PerturbativeDiagram}, 
the corresponding counterterm diagrams of Fig.~\ref{GluonicCountertermDiagrams}, 
and  the renormalized diagram are given in Table~\ref{glue_mix_diag_renorm_tab}.

\begin{table}[htb]
    \centering
    \begingroup
    \renewcommand{\arraystretch}{1.5}
    \begin{tabular}{?c?c?c?}
    \thickhline
    Topology & Gluon Vertex & Quark Vertex\\
    \hline
    Subdivergence & $4m_q \frac{\alpha}{\pi} \frac{1}{\epsilon}$  & $0$ \\
    \thickhline
    \end{tabular}
    \endgroup 
    \caption{
    Gauge-independent divergent parts of the subdiagrams of Fig.~\ref{GluonicSubdiagrams}  
    (i.e., subdiagrams originating from  Fig.~\ref{Glue_mix_PerturbativeDiagram}). 
    }
    \label{glue_mix_subdiag_tab}
\end{table}

\begin{table}[htb]
    \centering
    \begingroup
    \renewcommand{\arraystretch}{1.5}
    \begin{tabular}{?c?c?c?c?c?}
    \thickhline
 Diagram & Bare & \renewcommand{\arraystretch}{0.95}\begin{tabular}{@{}c@{}}Gluon Vertex \\ Counterterm\end{tabular}  & \renewcommand{\arraystretch}{0.95}\begin{tabular}{@{}c@{}}Quark Vertex \\ Counterterm\end{tabular} & Renormalized \\
    \hline
$A$ &  $\frac{3}{\epsilon}-\frac{35}{2}$ & $\frac{3}{\epsilon}-6$ &  $0$ &
$-\frac{23}{2}$
  \\
  \hline
  $B$ & $3$ &  $\frac{3}{2}$ &  $0$ & $\frac{3}{2}$ \\
    \thickhline
    \end{tabular}
    \endgroup 
    \caption{Diagrammatic renormalization contributions to $\Pi_{gq}$ in~\eqref{pi_gq_form} for the bare
    Feynman diagram of Fig.~\ref{Glue_mix_PerturbativeDiagram} and counterterm diagrams of Fig.~\ref{GluonicCountertermDiagrams}.  
    The bare entries are repeated from Table~\ref{glue_mix_conv_tab}. 
    The renormalized entries are obtained by subtracting the counterterms from the bare result. 
    All contributions are gauge-independent.
    }
    \label{glue_mix_diag_renorm_tab}
\end{table}

The agreement between the total result for $\Pi_{gq}$ in conventional and diagrammatic renormalization
(compare Tables~\ref{glue_mix_conv_tab} and~\ref{glue_mix_diag_renorm_tab}) is an impressive illustration of
diagrammatic renormalization methods in situations where there is composite operator mixing in conventional
renormalization.  The diagrammatic method required no prior knowledge of the underlying composite operator renormalization in~\eqref{GG_renorm}.  

This benchmark example for the quark/glueball mixed correlation function can also be used to build conceptual understanding of the relationship between conventional and diagrammatic renormalization.  
It is easily seen that subtraction of the Table~\ref{glue_mix_subdiag_tab} subdivergence for the gluon vertex
topology is identical to the renormalization-induced diagram generated by the coefficient of $\bar q q$ 
in~\eqref{GG_renorm}, and hence, the diagrammatic counterterm diagram and the conventional
renormalization-induced term are operationally identical.  The analogy can be taken even further by
recognizing that the gluon-vertex subdiagram would be used in conventional renormalization to project 
out the $\bar q q$ operator's contribution in~\eqref{GG_renorm}.  
Furthermore, the
absence of a quark-vertex divergence in Table~\ref{glue_mix_subdiag_tab} illustrates that, 
in conventional renormalization, the $\bar q q$ operator does not mix with gluonic operators 
as illustrated in the multiplicative renormalization result~\eqref{scalar_curr_renorm} 
for the $\bar q q$ operator.

\section{Discussion}
\label{discussion_sec}

\noindent
The Section~\ref{quark_glue_mix_sec} analysis of the mixed correlation function of scalar glueball 
and quark meson operators provided a conceptual interpretation of the relationship  between diagrammatic 
and conventional renormalization methods.  
The conceptual connections between diagrammatic and conventional renormalization  are still present in the
other benchmark examples presented, but these connections are a bit more difficult to discern. 
For example, in conventional renormalization for Feynman gauge, the Fig.~\ref{MesonicSubdiagrams} gluon exchange subdiagram  for the light-quark vector correlator is used to find the conventional 
renormalization of the vector  composite operator~\eqref{vec_curr_renorm}.  
The Table~\ref{vec_SD_tab} divergence in the gluon exchange subdiagram  corresponds to the quark field 
renormalization for the external quark lines, 
which establishes that no additional renormalization factor is needed for the vector  composite operator
(i.e., Eq.~\eqref{vec_curr_renorm} corresponds to $Z=1$).  
The  Fig.~\ref{MesonicSubdiagrams} self-energy subdiagram also corresponds to the quark field renormalizations as reflected by the same $\frac{1}{3\epsilon}$ prefactors in  Table~\ref{vec_SD_tab}.  
When the self-energy counterterm diagram of Fig.~\ref{MesonicCountertermDiagrams} is calculated, the $\slashed{p}$ factor conspires to cancel one of the propagators so that the self-energy and gluon-exchange
counterterm diagrams have the same loop-integration structure and  cancel as seen from
Table~\ref{vec_diag_combined_tab}.  

A similar (but more subtle) conceptual interpretation applies to the scalar current case in Feynman gauge,
except that the Fig.~\ref{MesonicSubdiagrams} gluon exchange subdiagram divergence corresponds to a
combination of the quark field renormalization for the external quark lines and the renormalization factor 
in~\eqref{scalar_curr_renorm}. 
However, the Fig.~\ref{MesonicSubdiagrams} self-energy subdiagram is still related to the quark field
renormalizations, and hence, 
the prefactors of the subdivergences in Table~\ref{scalar_SD_tab} are now different.  
The self-energy subdivergence
 $\slashed{p}$ factor again conspires to cancel one of the propagators so that the Fig.~\ref{MesonicCountertermDiagrams} self-energy and gluon-exchange counterterm diagrams have the same structure as the Fig.~\ref{MesonicPerturbativeDiagrams} LO bare diagram.  
 However, because $Z\ne 1$ for the scalar operator renormalization, the two counterterms do not 
 cancel, as seen in Table~\ref{scalar_diag_combined_tab}, but instead, their residual
 contribution is from the scalar operator renormalization factor~\eqref{scalar_curr_renorm}.  
 From Table~\ref{scalar_diag_combined_tab}, the residual contribution from the counterterms is $4-\frac{2}{\epsilon}$, which when subtracted, is identical to the conventional renormalization 
 contribution (which is also proportional to the Fig.~\ref{MesonicPerturbativeDiagrams} LO bare diagram) in Table~\ref{scalar_conv_tab}.

Thus, the diagrammatic renormalization results (e.g., Tables~\ref{vec_diag_combined_tab} 
and~\ref{scalar_diag_combined_tab}) can be parsed in two ways.  
The bare and counterterm contributions for each diagram can be combined to cancel the divergence in each diagram leading to the total renormalized result emerging from the sum of the
renormalized diagrams.
Alternatively, the counterterm contributions can be combined to reconstruct the conventional renormalization-induced contribution (e.g., Table~\ref{scalar_conv_tab}). 

With the conceptual connections between diagrammatic and conventional renormalization now established, the advantages of the diagrammatic method become more apparent.  In both the conventional and diagrammatic approaches, similar subdiagrams need to be computed.  In the diagrammatic case, these subdiagrams naturally emerge from the underlying diagrams of the loop expansion, but, 
in the conventional method, one is calculating Green functions containing the composite operator and fundamental fields  with only the general guiding principles of composite operator renormalization. 
In the conventional approach, it may be necessary to carefully choose and analyze the Green functions  to project out and disentangle the mixings in the underlying composite operator's renormalization structure.  After the entire conventional operator renormalization process is complete, it is then necessary to go back and compute the new diagrams resulting from composite operator renormalization, including any operator mixing. By contrast, the diagrammatic method does not need to disentangle the structure of the subdiagrams and it does not matter if the underlying structure comes from multiple operators; 
all that is necessary is to include the subdiagram's divergence into the associated counterterm diagram.
Thus, diagrammatic renormalization methods provide a considerable increase in computational efficiency for QCD sum-rule correlation functions.

A consistency check exists at each stage of the diagrammatic approach because the non-local divergences of 
an individual diagram must be cancelled by the counterterms generated by the subdiagrams.  
By contrast, in the conventional approach, the consistency check 
comes only at the final stage of the calculation
where the non-local divergences must cancel in the total result summing all bare diagrams and all renormalization-induced operator mixing diagrams.  
In the situation where conventional renormalization may involve mixing of many operators 
(e.g., dimension-six four-quark operators of Refs.~\cite{Narison:1983kn}), 
it may be difficult to isolate calculation errors that could exist within any of the renormalization constants, individual renormalization-induced  diagrams, or individual bare diagrams.  

In summary, the motivation for this paper is the need to develop efficient renormalization techniques for 
NLO  QCD sum-rule analyses, what  
with numerous experimental discoveries of exotic hadrons such as  tetraquarks, pentaquarks and  hybrids 
(see e.g., Refs.~\cite{Brambilla:2019esw,Liu:2019zoy,Meyer:2015eta} for reviews).  
For these NLO QCD sum-rule analyses, it is necessary to  renormalize correlation functions of QCD composite operators of high mass dimension (e.g., $15/2$ for pentaquarks).  

The conventional renormalization of composite operators becomes increasingly complicated as mass dimension increases, possibly requiring a large basis of operators mixed under renormalization.  
In conventional renormalization, the calculation of a QCD correlation function is a 
two-step process of first renormalizing the composite operator, 
and second, calculating renormalization-induced diagrams arising from the operator mixing under renormalization.  Thus, there are strong motivations to develop more efficient composite operator renormalization methodologies for QCD sum-rule applications.

Diagrammatic renormalization methods provide a compelling alternative to conventional renormalization by obviating the need to disentangle operator mixing.  
QCD sum-rule examples for vector and scalar mesons 
(Sections~\ref{vec_corr_sec}, \ref{scalar_corr_sec}, \ref{heavy_quark_section}), 
diquarks (Section~\ref{diquark_renorm_section}), 
and scalar quark/glueball mixing (Section~\ref{quark_glue_mix_sec}) have been presented in both conventional and diagrammatic approaches to illustrate the validity and advantages of diagrammatic methods in QCD sum-rules, and to highlight subtleties in the diagrammatic method.    Key steps of the diagrammatic renormalization methodology in QCD sum-rules  are summarized in Section~\ref{appendix_sec}  along with  results for extracting divergent parts of subdiagrams.
It is hoped that the efficiency and internal self-consistency checks of diagrammatic methods will provide the necessary tools for QCD sum-rule practitioners to tackle the very challenging prospect of NLO QCD sum-rule 
analyses for exotic hadrons.

\section*{Acknowledgments}
\noindent
TGS and DH are grateful for research funding from the Natural Sciences and Engineering Research Council of Canada (NSERC).


\section{Appendix: Methodological Summary and Key Results}
\label{appendix_sec}

\noindent
In this Appendix, a methodological summary  is provided to guide the application of diagrammatic renormalization methods to QCD sum-rules for two-point correlation functions, with  a particular view to applications to NLO contributions in multi-quark systems such as tetraquarks and pentaquarks.    Key results for the divergent parts of one-loop integrals are also provided to support the outlined methodology.

The  diagrammatic renormalization methodology   begins by constructing all (NLO) loop diagrams for the desired  two-point  QCD sum-rule correlation function. Each individual diagram is then renormalized using the following steps.  
Each step in the process includes an action, possible consistency check(s), and items to note.

\clearpage
\begin{enumerate}
\item {\bf Calculate Bare Diagram }
\begin{description}
\item[Action:] Calculate the bare diagram (in dimensional regularization), discarding any local divergences (corresponding to dispersion relation subtractions), retaining all finite parts,  and paying particular attention to the non-local divergences.  

\item[Consistency Checks:] Perform calculations in arbitrary covariant gauge or two different gauges;  proceed through method even for diagrams  without non-local divergences. 

\item[Notes:]  Make initial choice of renormalization scheme (e.g.,  $\overline{\rm MS}$ scheme) and associated  renormalization scale.   
\end{description}

\item  {\bf Construct Subdiagrams and Counterterm Diagrams} 
\begin{description}
\item[Action:] For each bare diagram, identify  every subdiagram (i.e., containing one or more loops) and construct  the corresponding counterterm diagram.  Discard subdiagrams and counterterm diagrams (such as in Fig.~\ref{additional_subdiagram_fig}) that  do not carry the external momentum because they represent local divergences (corresponding to dispersion relation subtractions)

\item[Consistency Checks:] Perform calculations in arbitrary covariant gauge or two different gauges;  proceed through method even for subdiagrams emerging from bare diagrams  without non-local divergences. 

\item[Notes:] Ensure that Dirac and Lorentz structures internal to the subdiagram are isolated so that only  external line degrees of freedom remain to form counterterm diagram.

\end{description}

\item {\bf Calculate Divergent Parts of Subdiagrams (Subdivergences)}

\begin{description}

\item [Action:] Extract divergent part of the subdiagrams (see Eqs.~\eqref{2pt_1}--\eqref{3pt_k_k_k_k} for divergent parts of selected one-loop integrals) and ignore all finite parts.

\item[Consistency Checks:]  Bare diagrams without non-local  divergences should have finite subdiagrams; such verification should occur in an arbitrary covariant gauge or two different gauges.  

\item[Notes:] Ensure that Dirac and Lorentz structures internal to the subdiagram are evaluated and not deferred to later stages of calculation;  result should  only carry external line degrees of freedom. 
\end{description}

\item {\bf Calculate Counterterm Diagrams}
\begin{description}
\item[Action:] Use divergent parts of subdiagrams (subdivergences) to calculate counterterm diagrams, retaining finite parts, discarding any  local  contributions (corresponding to dispersion relation subtractions), and paying particular attention to non-local divergences.

\item[Consistency Checks:] Perform calculations in arbitrary covariant gauge or two different gauges.

\item [Notes:]  Align renormalization scale with scheme chosen for bare diagram  (e.g., $\overline{\rm MS}$ scheme) 

\end{description}

\item {\bf Calculate Renormalized Diagram}

\begin{description}

\item[Action:] Subtract  results for counterterm diagram(s) from bare diagram to find renormalized diagram.

\item[Consistency Checks:] Ensure cancellation of non-local divergences  in renormalized diagram; verify cancellation in    an arbitrary covariant gauge or in two different gauges.  

\item[Notes:] Make final decision on renormalization scheme, converting renormalization scale as needed (e.g., from   $\rm MS$  to $\overline{\rm MS}$ scheme) and interpret coupling and mass parameters in renormalized result as $\alpha(\nu)$, $m(\nu)$  for renormalization scale $\nu$ in the chosen scheme.
\end{description}

\end{enumerate}
\noindent
After all individual diagrams have been renormalized, the renormalized diagrams are combined to obtain the final result for the renormalized QCD sum-rule correlation function, with coupling and mass parameters interpreted as $\alpha(\nu)$, $m(\nu)$  for renormalization scale $\nu$ in the chosen scheme.  
The final result for the correlattion function should be gauge independent,  providing an additional consistency check  supplementing those outlined in the above summary. 
Anomalous dimensions for the correlation function can be extracted from the final result.

Divergent  parts of  one-loop Feynman integrals used in our examples and that are anticipated  to occur for one-loop subdiagram topologies in future NLO applications to QCD sum-rules (including multiquark systems) are now  outlined  for our dimensional regularization convention $D=4+2\epsilon$.   For two-point subdiagram topologies containing one external momentum scale in loop integrals (e.g.,  diagrams $a$ and $b$ in Fig.~\ref{HLSubdiagrams}), useful results for divergent parts are as follows:
\begin{gather}
\int \frac{d^Dk}{(2\pi)^D} \frac{1}{\left(k^2-m_1^2\right)\left((k+q)^2-m_2^2\right)}=-\frac{i}{16\pi^2\epsilon}+ O(\epsilon)
\label{2pt_1}
\\
\int \frac{d^Dk}{(2\pi)^D}\frac{k^\mu}{\left(k^2-m_1^2\right)\left((k+q)^2-m_2^2\right)}=\frac{i}{32\pi^2\epsilon}q^\mu+O(\epsilon)
\label{2pt_k}
\\
\int \frac{d^Dk}{(2\pi)^D} \frac{k^\mu k^\nu}{\left(k^2-m_1^2\right)\left((k+q)^2-m_2^2\right)}
=\frac{i}{48\pi^2\epsilon}\left( \frac{1}{4}q^2 g^{\mu\nu}-q^\mu q^\nu \right)+O(\epsilon)
\label{2pt_k_k}
\\
\int \frac{d^Dk}{(2\pi)^D} \frac{k^\mu k^\nu}{k^4\left((k-q)^2-m_2^2\right)}
=-\frac{i}{64\pi^2\epsilon}g^{\mu\nu}+O(\epsilon)
\label{2pt_landau_k_k}
\\
\int \frac{d^Dk}{(2\pi)^D} \frac{k^\rho k^\lambda k^\omega}{k^4\left((k-q)^2-m_2^2\right)}
=-\frac{i}{192\pi^2\epsilon}\left(g^{\rho\omega} q^\lambda + g^{\rho\lambda}q^\omega+ g^{\lambda\omega} q^\rho   \right)+O(\epsilon)\,,
\label{2pt_landau_k_k_k}
\end{gather}
where the divergent parts are independent of the propagator masses $m_1$, $m_2$ and Eqs.~\eqref{2pt_landau_k_k} and \eqref{2pt_landau_k_k_k}  emerge from gauge parameter dependence and are therefore written for $m_1=0$ to emphasize this point. Similarly, for  three-point subdiagram topologies containing two external momentum scales in loop integrals 
(e.g.,  Fig.~\ref{GluonicSubdiagrams}), useful results for divergent parts are as follows:
\begin{gather}
\int \frac{d^Dk}{(2\pi)^D} \frac{1}{\left(k^2-m_1^2\right)\left((k-q)^2-m_2^2\right)\left((k-p)^2-m_3^2\right)}=0+O(\epsilon)
\label{3pt_1}
\\
\int \frac{d^Dk}{(2\pi)^D} \frac{k^\mu}{\left(k^2-m_1^2\right)\left((k-q)^2-m_2^2\right)\left((k-p)^2-m_3^2\right)}=0+O(\epsilon)
\label{3pt_k}
\\
\int \frac{d^Dk}{(2\pi)^D} \frac{k^\mu k^\nu}{\left(k^2-m_1^2\right)\left((k-q)^2-m_2^2\right)\left((k-p)^2-m_3^2\right)}=-\frac{i}{64\pi^2\epsilon}g^{\mu\nu}+O(\epsilon)
\label{3pt_k_k}
\\[5pt]
\begin{split}
\int \frac{d^Dk}{(2\pi)^D} &\frac{k^\rho k^\lambda k^\omega}{\left(k^2-m_1^2\right)\left((k-q)^2-m_2^2\right)\left((k-p)^2-m_3^2\right)}
\\
&\qquad\qquad\qquad=-\frac{i}{192\pi^2\epsilon}\left(g^{\rho\omega}(p+q)^\lambda + g^{\rho\lambda}(p+q)^\omega+ g^{\lambda\omega}(p+q)^\rho   \right) +O(\epsilon)
\label{3pt_k_k_k}
\end{split}
\\[5pt]
\begin{split}
&\int \frac{d^Dk}{(2\pi)^D} \frac{k^\rho k^\lambda k^\mu k^\nu}{\left(k^2-m_1^2\right)\left((k-q)^2-m_2^2\right)\left((k-p)^2-m_3^2\right)}
\\
&~=-\frac{i}{768\pi^2\epsilon}\left( g^{\mu\nu}\left( 2p^\lambda p^\rho +2q^\lambda q^\rho+ p^\lambda q^\rho + q^\lambda p^\rho\right)  
+g^{\mu\nu}g^{\lambda\rho}\left(p^2+q^2-p\cdot q\right)
  + {\rm permutations}
\right)+O(\epsilon)
\end{split}
\label{3pt_k_k_k_k}
\end{gather}
where  the permutations in \eqref{3pt_k_k_k_k} indicate those necessary to form a completely symmetric tensor in $\{\mu,\nu,\lambda,\rho\}$ and zero in Eqs.~\eqref{3pt_1} and \eqref{3pt_k} indicate that the integral is finite.

\end{document}